\documentstyle[12pt]{article}
\begin{document}
\title{
\begin{flushright}
{\normalsize Gunma-Tech-94-1}
\end{flushright}
\vspace{1cm}
Gap Equations of $O(N)$ Non-linear $\sigma$ Model in Three
Dimensions}
\author{Kazuto Oshima
\\ \\
\sl Gunma College of Technology, Maebashi 371}
\date{  }
\maketitle
\\ \\
\begin{abstract}
We study the $O(N)$ non-linear $\sigma$ model on three-dimensional
manifolds of constant curvature by means of the large $N$ expansion
at the critical point. We examine saddle point equations imposing
anti-periodic boundary condition in time direction.  In the case
$S^1 \times S^2$ we find that a solution is inevitably unstable.
We briefly refer to the case $S^1 \times S^1 \times S^1$.
\end{abstract}
%
%
\section{Introduction}
\qquad Two-dimensional conformal field theories (CFT) \cite{BPZ}
have greatly been
developed in the last decade.   Various important findings
have been obtained.
On the contrary, much yet remains to be investigated in
higher-dimensional
CFT \cite{car1}. It will be especially important to study three-dimensional CFT
in the context of understanding critical behavior of three-dimensional
statistical models. It is known that the three-dimensional   $O(N)$
non-linear
$\sigma$ model \cite{are,guru} becomes conformally invariant at
a non-trivial fixed point.
It is worth while studying this model as a three-dimensional CFT.

In a slab with periodic boundary condition (PBC), a solution of saddle
point equations in the three-dimensional $O(N)$ non-linear $\sigma$
model is possessed of a remarkable feature. It has been shown by
Sachdev \{sach} that a universal number, which is called ''central charge''
and is calculated from the finite size scaling of the free energy \cite{fis},
is
given by a simple rational number.  Critical properties
of this model in various manifold of constant curvature with PBC have been
investigated by Guruswamy et. al. \cite{guru} in terms of the zeta function
regularization method.

In two-dimensional CFT, it is well known that modular properties play
an important role \cite{cap}.   Cardy \cite{car2} has proposed a possible
generalization
of modular invariance in higher-dimensional CFT. In time direction of
$S^1$, anti-periodic boundary condition (APBC) should be imposed to
obtain a modular invariant partition function in odd dimensions.
A gap equation and modular invariant partition function of the $O(N)$
non-linear $\sigma$ model on $S^1 \times S^2$ with APBC have already
been studied by Fujii and Inami \cite{fuji}.  It is difficult to solve the
gap equation exactly and deviation from $S^1 \times {\bf R}^2$ or
${\bf R} \times S^2$ has been estimated in certain limit. The purpose of
this paper is to examine gap equations more subtly.  We take possible
spontaneous symmetry breaking (a non-vanishing vacuum expectation value of
a scalar field) into account, which makes it easy to compare the case
of APBC with that of PBC.  We discuss whether a solution is stable
or not. If the solution is unstable, the corresponding partition
function is supposed to be out of meaning.

In sect.2, we introduce the $O(N)$ non-linear $\sigma$ model and gap
equations.  In sect.3, we examine the case $S^1 \times {\bf R}^2$ for
later use. In sect.4, we study the gap equations on $S^1 \times S^2$.
In sect.5, the case of $S^1 \times S^1 \times S^1$ is briefly treated.
The last section is devoted to summary.
\section{$O(N)$ non-linear $\sigma$ model and gap equations}
\qquad In this section, we introduce the $O(N)$ non-linear $\sigma$ model
on a three-dimensional Riemannian manifold $({\sl M},g)$. Our main interest
is the gap equations in the large $N$ limit.

The euclidean partition function of the $O(N)$ non-linear $\sigma$
model on the three-dimensional manifold ${\sl M}$ is given by
\begin{equation}
Z[g,\lambda]=\int D[\phi]\exp\Bigl(-{1 \over 2\lambda}\int_{\sl M}
d^{3}x \sqrt{g} g^{\mu \nu}(x)\partial_{\mu} \phi^{i}
\partial_{\nu}\phi_{i} \Bigr),
\end{equation}
where $i=1,2\ldots N $, $\lambda$ is a coupling constant and the
scalar fields $\phi^{i}(x)$ satisfy the constraint $\phi^{i}(x)
\phi_{i}(x)=1$. The action is not invariant under the conformal
transformation of the metric, $g^{\mu\nu}(x)\longrightarrow
\Omega^{2}(x)g^{\mu\nu}(x)$. We modify the action as follows:
\begin{equation}
S[g,\lambda]={1 \over 2\lambda}\int_{\sl M} d^{3}x \sqrt{g}\mbox{[}
-\phi^i\Box_{g}\phi_{i}\mbox{]},
\end{equation}
where $\Box_{g}=-{1 \over \sqrt{g}}\partial_{\mu}(\sqrt{g}
g^{\mu\nu}\partial_{\nu})+\xi R$ is the conformal laplacian.
The number $\xi=(d-2)/4(d-1)$ is equal to $1/8$ in three dimensions
and $R$ denotes the Ricci scalar curvature.   The action becomes
conformally invariant \cite{bir} under the transformation with $\phi^{i}$
transforming as $\phi^{i}(x)\longrightarrow\Omega^{1-\frac{d}{2}}
(x)\phi(x)$. Owing to the constraint $\phi^{i}(x)\phi_{i}(x)=1$,
classically conformal invaraiance is still violated. But it is
known that there is a non-trivial fixed point for the quantum
theory at which it is conformally invariant.

Introducing an auxiliary field $\sigma(x)$, the modified partition
function is expressed as
\begin{equation}
Z[g,\lambda]=\int D[\phi] D[\sigma]\exp\Bigl(-{1 \over 2\lambda}
\int_{\sl M}d^{3}x \sqrt{g}\mbox{[} -\phi^{i}\Box_{g}\phi_{i}
+\sigma(\phi^{i}\phi_{i}-1)}\mbox{]} \Bigr).
\end{equation}
We redefine $(N-1)\lambda$ as $\lambda$ which will remain fixed as
$N \rightarrow \infty.$ We also redefine $\phi^{i}$ as $\sqrt{\lambda}
\phi^{i}.$  Carring out the path integration for the first $N-1$
components of the fields $\phi^{i}$, the partition function is now
formally be written as
\begin{equation}
Z[g,\lambda]=\int D[\phi_{N}] D[\sigma]\exp\Bigl(-{N-1 \over 2}
\Bigl[TrLog(-\Box_{g}+\sigma)+\int_{\sl M}d^{3}x \sqrt{g}
\mbox{[}\phi^{N}(-\Box_{g}+\sigma)\phi_{N}-
\frac{\sigma}{\lambda}\mbox{]}\Bigl] \Bigr).
\end{equation}
Here, using the $O(N)$ symmetry, we have chosen $\phi_{N}$ as a field
which may have a non-vanishing vacuum expectation value.  We seek a uniform
saddle point solution of the form $\sigma(x)=m^{2}, \phi_{N}(x)=b$.
This is given by the following gap equations:
\begin{equation}
(-\Box_{g}+m^{2})b=0,
\end{equation}
\begin{equation}
b^{2}= \lim_{s \to 1} \bigl(\frac{1}{\lambda (s)}
-<x|(-\Box_{g}+m^{2})^{-s}|x> \bigr),
\end{equation}
where the zeta function regularization method is meant
on the right hand side of Eq.(6).
The two parameters $m$ and $b$ are the physical mass and
the vacuum expectation value
of the $\phi_{N}$ field respectively. Although, APBC will imply
$b=0$ physically, we do not set so at the start for convenience of
comparison with the PBC case. At the saddle point, the free energy
density ${\cal F} [ g,\lambda]$ is given by
\begin{equation}
\frac{{\cal F }[ g,\lambda]}{N}={1 \over 2} \Bigl(-\[ \lim_{s \to 0}
{d \over ds}<x|(-\Box_{g}+m^{2})^{-s}|x>
-\lim_{s \to 1}{ m^{2} \over \lambda(s)}\Bigr).
\end{equation}

It has been  discussed in Ref.4) that the divergent part of the propagator
is independent of the metric $g^{\mu \nu}(x)$ and the critical value
of the coupling constant is independent of the metric. Thus the
critical point $\lambda_{c}$ for a curved space is equal to the one
for the flat space ${\bf R}^3}$.  It has also been shown that
$\displaystyle \lim_{s \to 1} 1/\lambda_{c}(s)=0$ for the flat space
${\bf R}^{3}$. At the critical point, the second gap equation and
the free energy density are simplified as follows:
\begin{equation}
b^{2}= -\lim_{s \to 1} <x|(-\Box_{g}+m^{2})^{-s}|x> ,
\end{equation}
\begin{equation}
\frac{{\cal F}_{c}[ g]}{N}=-{1 \over 2} \[ \lim_{s \to 0}{d \over ds}
<x|(-\Box_{g}+m^{2})^{-s}|x>.
\end{equation}
We regard Eqs.(5),(8) and (9) as our starting point.
%
%
\section{Finite size effects on $ S^{1} \times {\bf R}^{2}$}
\qquad Before studying the gap equations on the manifold $S^{1} \times S^{2}$,
we examine the model on a slab $S^{1} \times {\bf R}^{2}$ for later
convenience.

\subsection{Case of periodic boundary condition}

This case has been well studied by Sachdev \cite{sach}.
Guruswamy et.al. \cite{guru}
have reproduced the same results in the zeta function
regularization method. On $S^{1} \times {\bf R}^{2}$ the gap
equations are
\begin{equation}
m^{2}b=0,
\end{equation}
\begin{equation}
b^{2}=- \lim_{s \to 1} {1 \over L}\sum_{n=-\infty}^{\infty}
\int {d^{2}k \over (2\pi)^{2}} \bigl( k^{2}+{4\pi^{2}n^{2} \over L^{2}}
+ m^{2} \bigr)^{-s},
\end{equation}
where $L$ is the radius of $S^{1}$. In the zeta function regularization
method, Eq.(11) is calculated as
\begin{eqnarray}
b^{2} &= &- \lim_{s \to 1} {1 \over L}\sum_{n=-\infty}^{\infty}
\int {d^{2}k \over (2\pi)^{2}}\int_{0}^{\infty}dt{t^{s-1} \over \Gamma(s)}
e^{-t \bigl( k^{2}+{4\pi^{2}n^{2} \over L^{2}}+ m^{2} \bigr)} \nonumber \\
 &= &- \lim_{s \to 1} {1 \over \sqrt{4\pi}} \int_{0}^{\infty}dt
{t^{s-{3 \over2}} \over \Gamma (s)}
\int {d^{2}k \over (2\pi)^{2}}\sum_{n=-\infty}^{\infty}
e^{- \bigl(t k^{2}+{L^{2}n^{2} \over 4t}+t m^{2} \bigr)} \nonumber \\
 &= &{m \over 4\pi}-{1 \over 2\pi L} \sum_{n=1}^{\infty}
{e^{-mLn} \over n} \nonumber \\
 &= &{1 \over 2\pi L}\ln2\sinh{mL \over 2},
\end{eqnarray}
where we have used the Poisson summation formula:
\begin{equation}
\sum_{n=-\infty}^{\infty} e^{-{4 \pi^{2} n^{2} \over L^{2}}t}
={L \over \sqrt{4\pi t}}\sum_{n=-\infty}^{\infty}e^{-{L^{2}n^{2} \over 4t}}.
\end{equation}
{}From Eqs.(10) and (12), we find $b=0, m=m_{L}=2\ln \tau/L$,
where $\tau$ is the Golden mean ${1+\sqrt{5} \over 2}$.  For this
solution $m=m_{L}$, the free energy density is computed as
\begin{eqnarray}
{{\cal F}_{c} \over N} &=&- \lim_{s \to 0}{d \over ds}
 {1 \over 2L}\sum_{n=-\infty}^{\infty}
\int {d^{2}k \over (2\pi)^{2}} \bigl( k^{2}+{4\pi^{2}n^{2} \over L^{2}}
+ m_{L}^{2} \bigr)^{-s} \nonumber \\
 &=& - \lim_{s \to 0} {1 \over 2L}\sum_{n=-\infty}^{\infty}
\int_{0}^{\infty} dt
\quad{t^{s-1} \over \Gamma (s)}\int {d^{2}k \over (2\pi)^{2}}
e^{-t \bigl( k^{2}+{4\pi^{2}n^{2} \over L^{2}}
+ m_{L}^{2} \bigr)} \nonumber \\
 &=& -{1 \over 2L} {d \over ds}\zeta (0),
\end{eqnarray}
where $\zeta (s)$ is given by
\begin{eqnarray}
\zeta (s) &=& {1 \over \Gamma (s)}\sum_{n=-\infty}^{\infty}
\int_{0}^{\infty} dt\quad t^{s-1} \int {d^{2}k \over (2\pi)^{2}}
e^{-t \bigl( k^{2}+{4\pi^{2}n^{2} \over L^{2}}+ m_{L}^{2} \bigr)} \nonumber \\
 &= & {L \over \sqrt{4\pi}\Gamma (s)}\sum_{n=-\infty}^{\infty}
 \int_{0}^{\infty}dt\quad t^{s-{3 \over2}}
\int {d^{2}k \over (2\pi)^{2}}
e^{- \bigl(t k^{2}+{L^{2}n^{2} \over 4t}+t m_{L}^{2} \bigr)} \nonumber \\
 &= & {L \over ({4\pi})^{3 \over 2}\Gamma (s)}\sum_{n=-\infty}^{\infty}
 \int_{0}^{\infty}dt\quad t^{s-{5 \over2}}
e^{- {L^{2}n^{2} \over 4t}-t m_{L}^{2} } \nonumber \\
 &=& {L m_{L}^{3-2s} \over (4 \pi)^{3 \over 2} \Gamma(s)}
\Bigl( \Gamma(s-{3 \over 2})+ 2\sum_{n \neq 0}
\Bigl({|n|Lm_{L} \over 2}\Bigr)^{s-{3 \over 2}} K_{{3 \over 2}-s}
(|n|Lm_{L})\Bigr).
\end{eqnarray}
Here, $K_{\nu}$ is the modified Bessel function:
\begin{equation}
K_{\nu}(z)= {1 \over 2}\Bigl({z \over 2}\Bigr)^{\nu}\int_{0}^{\infty}
dt\quad t^{-\nu-{1 \over 2}} e^{-t-{z^{2} \over 4t}},
\end{equation}
for  Re$\nu > -1/2$,$|$arg $z|< \pi/4$. In the limit $s \rightarrow 0$,
$1 / \Gamma (s)$ behaves as $s$ and $\zeta^{\prime} (0)$ is
found to be
\begin{equation}
\zeta^{\prime} (0)={L m_{L}^{3} \over (4\pi)^{3 \over 2}}
(\Gamma(-{3 \over 2})+4\sum_{n=1}^{\infty}\Bigr({2 \over nL m_{L}}\Bigr)
^{3 \over 2} K_{3 \over 2}(nL m_{L})).
\end{equation}
The free energy density is calculated to be
\begin{equation}
{{\cal F}_{c} \over N}=-{m_{L}^{3} \over 2\pi}({1 \over 6}
+\sum_{n=1}^{\infty}\Bigl( {e^{-nL m_{L}} \over (nL m_{L})^{3}}
+{e^{-nL m_{L}} \over (nL m_{L})^{2}} \Bigr)).
\end{equation}
The right hand side of this equation is expressed by the polylogarithmic
function $Li_{p}(z)=\sum_{n=1}^{\infty} z^{n}/ n^{p}}$ and is
identical to the following simple expression \cite{sach}:
\begin{equation}
{{\cal F}_{c} \over N}=-{\zeta _{R}(3) \over 2\pi L^{3}}\cdot
{4 \over 5},
\end{equation}
where $\zeta_{R}$ is the Riemann zeta function. Comparing this with
the expression for the free energy density obtained from hyperscaling
for a d-dimensional slab \cite{fis}:
\begin{equation}
{\cal F}={\cal F}_{\infty}-{\Gamma({d \over 2}) \zeta_{R} (d)
\over \pi^{d \over 2} L^{d}}\tilde{c},
\end{equation}
we see $\tilde{c}/N =4/5$. Here, $\tilde{c}$ is a universal number.
In two-dimensions, it coincides with the central charge in CFT.

\subsection{Case of anti periodic-boundary condition}
In this case, the second gap equation is given by Eq.(11) with $n$
replaced by $n+{1 \over 2}$. Using the relevant Poisson summation formula
\begin{equation}
\sum_{n=-\infty}^{\infty} e^{-{4\pi^{2}(n+1/2)^{2} \over L^{2}}t}
={L \over \sqrt{4\pi t}}\Bigl( 1+2\sum_{n=1}^{\infty} (-1)^{n}
e^{-{L^{2}n^{2} \over 4t}}\Bigr),
\end{equation}
we get
\begin{eqnarray}
b^{2} &= & \lim_{s \rightarrow 1}
{1 \over ({4\pi})^{3 \over 2}\Gamma (s)}
 \int_{0}^{\infty}dt\quad t^{s-{5 \over2}}(e^{-t m^{{2}}}+2\sum_{n=1}^{\infty}
e^{- {L^{2}n^{2} \over 4t}-t m^{2} }) \nonumber \\
 &= & -{m \over (4\pi )^{3 \over 2}}\Gamma (-1/2)
-{1 \over {2\pi L}}\sum_{n=1}^{\infty}(-1)^{n}{e^{-mLn} \over n} \nonumber \\
 &= & {1 \over {2\pi L}} \ln2 \cosh{mL \over 2}.
\end{eqnarray}
The first gap equation (10) leads to $b=0$ or $m=0$. Setting $b=0$ in
Eq.(22),$ m$ is given by an imaginary number.  This solution corresponds
to a tachyonic mode and is unstable. The other possibility is
$m=0, b^{2}=\ln 2 /2\pi L$. The free energy density which corresponds
to the latter solution is
\begin{eqnarray}
{{\cal F}_{c} \over N} &=& -\lim_{s \rightarrow 0}{d \over ds}
\lim_{m \rightarrow 0}{1 \over 2L} \sum_{n=-\infty}^{\infty}
\int {d^{2}k \over (2\pi )^{2}}(k^{2}+{4\pi ^{2}(n+1/2)^{2} \over L^{2}}
+m^{2})^{-s}  \nonumber \\
 &=& -\lim_{s \rightarrow 0}{1 \over (4\pi )^{3 \over 2}\Gamma (s)}
\sum_{n=1}^{\infty}(-1)^{n} \int_{0}^{\infty}dt \;t^{s-{5 \over 2}}
e^{-{L^{2}n^{2} \over 4t}} \nonumber \\
 &=& -{1 \over 2\pi L^{3}}\sum_{n=1}^{\infty}{(-1)^{n} \over n^{3}}
\nonumber \\
 &=& -{\zeta _{R}(3) \over 2\pi L^{3}}\cdot (-{3 \over 4}).
\end{eqnarray}
Comparring this expression with the general one of the finite size scaling,
we see
\begin{equation}
{\tilde{c} \over N}=-{3 \over 4}.
\end{equation}
Making use of the two-point function of the energy-momentum tensor
$T_{\mu \nu}$, Cardy \cite{car3} has proposed a possible definition of
the ``central charge'' $c$ in higher-dimensions.  According to Ref.5),
$c/N$ is computed to be $3/4$ for the $O(N)$ non-linear $\sigma$ model.
So, up to the negative sign,$ \tilde{c}$ coincides with $c$ for the case
of APBC. Althogh a non-vanishing value of $b$ may conflict with APBC,
there may be some meaning in the latter solution.
%

\section{Gap equations on $S^{1} \times S^{2}$}
\qquad In this section, we investigate gap equations on $S^{1} \times S^{2}$.
It will be difficult to seek a solution for finite radii. So, we examine
some limited cases.
\subsection{Deviation from $S^{1} \times {\bf R}^{2}$}

In this and in the following subsections, $\rho$ the radius of $S^{2}$
is considered to be finite but sufficiently large. The second gap
equation is
\begin{eqnarray}
-4\pi b^{2} &=& {1 \over L\rho ^{2}}\lim_{s \rightarrow 1}
\sum_{l= 1/2}^{\infty}\sum_{n=-\infty}^{\infty} 2l
(({l \over \rho})^{2} +({2\pi n \over L})^{2}+m^{2})^{-s} \nonumber \\
 &=& {1 \over L}\sum_{n=-\infty}^{\infty}\lim_{s \rightarrow 1}
\int_{0}^{\infty}dt{t^{s-1} \over \Gamma (s)}{1 \over \rho}
\sum_{l=1/2}^{\infty}2{l \over \rho}e^{-t(({l \over \rho})^{2}
+({2\pi n \over L})^{2}+m^{2})},
\end{eqnarray}
where $l$ runs over positive half-integers. To evaluate the right hand
side of Eq.(25), we use the Euler-Maclaurin formula as in Ref.9);
\begin{equation}
{1 \over L}\sum_{n}f({n \over L})=\int_{-\infty}^{\infty}dxf(x)
+(f^{\prime}(\infty)-f^{\prime}(-\infty))B_{\pm}+C_{\pm}(L),
\end{equation}
where the plus(minus) sign denotes that $n$ runs over
integers(half-integers),
$B_{+}=-{1 \over 12L^{2}},B_{-}={1 \over 24L^{2}},C_{+}(L)$ is an
$O(1 / L^{3})$ term and $C_{-}(L)$ is an $O(1 / L^{4})$
term. Applying Eq.(26) to $f(x)=2xe^{-t(x^{2}+({2\pi n \over L})^{2}
+m^{2})}$, Eq.(25) is calculated as
\begin{eqnarray}
-4\pi b^{2} &=& {1 \over L}\sum_{n=-\infty}^{\infty}\lim_{s \rightarrow 1}
\int_{0}^{\infty}dt{t^{s-2} \over \Gamma (s)}
e^{-t(({2\pi n \over L})^{2}+m^{2})}
+{1 \over 12L\rho ^{2}}\sum_{n=-\infty}^{\infty}\int_{0}^{\infty}dt
e^{-t(({2\pi n \over L})^{2}+m^{2})}+O(1 / \rho ^{4}) \nonumber \\
 &=& -{2 \over L}\ln2\sinh{mL \over 2}+{1 \over 12m\rho^{2}}\cosh{mL \over
2}+O(1 / \rho ^{4}).
\end{eqnarray}
Here, we have used the Poisson summation formula Eq.(13) in the course
of calculation. The first approximation of $m$ is given by $m_{L}(\ne 0)$.
So, from the first gap equation, we see $b=0$. Putting this value
into Eq.(27), we get
\begin{equation}
m=m_{L}+{L \over 48\rho ^{2}\sinh^{-1}{1 \over 2}}+O(1/\rho ^{4}).
\end{equation}
A similar result has already been obtained in Ref.9).

\subsection{Deviation from $S^{1} \times {\bf R}^{2}$ under APBC}

In this subsection, we impose APBC in the direction of $S^{1}$.
The second gap equation is given by Eq.(25) with $n$ replaced by
$n+{1 \over 2}$. Using the Poisson summation formula Eq.(21), we get
\begin{eqnarray}
-4\pi b^{2} &=& -{2 \over L}\ln 2\cosh{mL \over 2}+{\sinh{mL \over 2}
 \over 24m\rho ^{2}\cosh{mL \over 2}} +O(1/\rho^{3}) \nonumber \\
 &=& -{2 \over L}\ln 2\cosh{mL \over 2}+{L \over 48\rho ^{2}} +o(1/\rho^{3}).
\end{eqnarray}
Here, we have substituted $m=0$,which is one of the solutions in the
$\rho \rightarrow \infty$ limit, in the $O(1 /\rho ^{2})$ term.
Assuming that $b^{2}=\ln 2 /2\pi L$ still holds, $m$ is evaluated as
\begin{equation}
m={1 \over \sqrt{12} \rho} +O({1 \over \rho ^{2}}).
\end{equation}
A similar result can be seen in Ref.9). However, for a finite $\rho$
the Ricci scalar curvature is $R={2 \over \rho ^{2}}$ and the first gap
equation is given by
$(m^{2}+{1 \over 4\rho ^{2}})b=0$. So, for not such a large
$\rho $, we are obliged to choose the other solution $b=0$, and from Eq.(29),
 $m$ is given
by an imaginary number. Accordingly, it is supposed that a solution of the gap
equations on $S^{1} \times S^{2}$ is not stable.

There would be some uncertainty in the above discussion because of the
presence of the two solutions on $S^{1} \times {\bf R}^{2}$ with APBC.
There is a unique solution for the gap equations on ${\bf R} \times
S^{2}$. There will be no room for uncertainty in investigation
based on the analysis of ${\bf R} \times S^{2}$. In the following
subsection, we examine deviation from ${\bf R} \times S^{2}$.

\subsection{Deviation from ${\bf R} \times S^{2}$}

In this subsection, we consider $L$ the radius of $S^{1}$ to be finite but
sufficiently large. The second gap equation for $S^{1} \times S^{2}$
is
\begin{equation}
-4\pi b^{2}=\lim_{s \rightarrow 1}{1 \over \rho ^{2}}\sum_{l=1/2}^{\infty}
2l \int_{0}^{\infty}dt\quad {t^{s-1} \over \Gamma (s)}{1 \over L}
\sum_{n} e^{-t(({l \over \rho})^{2}+({2\pi n \over L})^{2}+m^{2})},
\end{equation}
where $n=$integer for PBC and $n=$half-integer for APBC. We use the
Euler-Maclaurin formula Eq.(26) for the summation over $n$. For $f(x)
=\exp (-x^{2})$, the second term in Eq.(26) vanishes and a  numerical
amalysis shows us that $C_{+}(L)>0$ and $ C_{-}(L)<0.$
The integral $\int_{-\infty}^{\infty}dx\exp (-x^{2})$ underestimates
(overestimates) the summation ${1 \over L}\displaystyle
\sum_{n=-\infty}^{\infty}
\exp (-n^{2}/L^{2})$ for PBC(APBC).
We will see in later that
negative value of $C_{-}$ causes unstability of a solution for APBC.
For $\exp (-t(2\pi n / L)^{2}), C_{+}$ is an  $O(t^{3 \over 2}
 / L^{4})$ term and $C_{-}$ is an $O(t / L^{3})$ term.
Taking the factor $t^{3 \over 2}(t)$ outside of $C_{+}(C_{-})$,
we get
\begin{eqnarray}
-4\pi b^{2} &=& \lim_{s \rightarrow 1}{1 \over 2\pi \rho^{2}\Gamma (s)}
\sum_{l=1/2}^{\infty}2l\int_{0}^{\infty}dte^{-t(({l \over \rho} )^{2}
+m^{2})}(t^{s-{3 \over 2}}+t^{s-{1 \over 2}}2\pi C_{+}),\quad {\rm (PBC)} \\
-4\pi b^{2} &=& \lim_{s \rightarrow 1}{1 \over 2\pi \rho^{2}\Gamma (s)}
\sum_{l=1/2}^{\infty}2l\int_{0}^{\infty}dte^{-t(({l \over \rho} )^{2}
+m^{2})}(t^{s-{3 \over 2}}+t^{s}2\pi C_{-}). \quad {\rm (APBC)}
\end{eqnarray}
The following type of the summation formula$^{4)}$ is useful:
\begin{equation}
{1 \over 2\pi}\sum_{l=1/2}^{\infty}2le^{-{l^{2} \over \rho^{2}}t}
={\rho^{2} \over 4\pi t}+{\rho^{2} \over (4\pi t)^{3 \over 2}}
{\rm P}\int_{-\infty}^{\infty}dx({x \over 2\rho}
{\rm cosec}{x \over 2\rho}-1)e^{-{x^{2} \over 4t}},
\end{equation}
where P denotes the principal value of the integral.
After changing variables, we obtain
\begin{eqnarray}
-4\pi b^{2} &=& -\sqrt{4\pi }m+{\sqrt{\pi}C_{+} \over 4m^{3}}
+{m \over 2\pi ^{3 \over 2}}{\rm P} \int_{-\infty}^{\infty}dx{1 \over x}
({1 \over 2m\rho}{\rm cosec}{x \over 2m\rho}-{1 \over x})|x|K_{1}(|x|)
\nonumber \\
 &+& { C_{+} \over 4\pi^{1 \over 2}m^{3}}{\rm P}\int_{-\infty}^{\infty}
dx{1 \over x}({1 \over 2m\rho}{\rm cosec}{x \over 2m\rho}-{1 \over x})
|x|^{3}K_{1}(|x|), \quad  {\rm (PBC)} \\
-4\pi b^{2} &=&  -\sqrt{4\pi }m+{C_{-} \over 2m^{2}}
+{m \over 2\pi ^{3 \over 2}}{\rm P} \int_{-\infty}^{\infty}dx{1 \over x}
({1 \over 2m\rho}{\rm cosec}{x \over 2m\rho}-{1 \over x})|x|K_{1}(|x|)
\nonumber \\
 &+& { C_{-} \over 4m^{2}}{\rm P}\int_{-\infty}^{\infty}
dx{1 \over x}({1 \over 2m\rho}{\rm cosec}{x \over 2m\rho}-{1 \over x})
x^{2}e^{-|x|}. \qquad     {\rm (APBC)}
\end{eqnarray}
Using the integral representation $K_{1}(x)={1 \over 2}\int_{0}^{\infty}
dt \exp (-({t \over 2}+{1 \over 2t})x)$, it is easy to see the integrals
$\int_{-\infty}^{\infty}dx |x|K_{1}(|x|)$ and $ \int_{-\infty}^{\infty}dx
|x|^{3}K_{1}(|x|)$ are finite. Noting that
\begin{equation}
{\rm P} \int_{-\infty}^{\infty}dx{1 \over x}({1 \over 2m\rho}
{\rm cosec}{x \over 2m\rho}-{1 \over x})=0,
\end{equation}
and that the integrands in Eqs.(35) and (36) oscillate rapidly for a small
$m\rho$(for $L=\infty, m=0$ and for a sufficiently large $L,\>m$ is
sufficiently small), it is supposed that the third and fourth terms
in Eqs.(35) and
(36) can be neglected. Thus we obtain
\begin{eqnarray}
-4\pi b^{2}&=&-\sqrt{4\pi}m+{\sqrt{\pi}C_{+} \over 4m^{3}},\qquad
{\rm (PBC)} \\
-4\pi b^{2}&=&-\sqrt{4\pi}m+{C_{-} \over 2m^{2}}. \quad \qquad {\rm (APBC)}
\end{eqnarray}
{}From the first gap equation $(m^{2}+{1 \over 4\rho^{2}})b=0$, we see
$b=0$.  The mass $m$ is computed as $m=(C_{+}/8)^{1/4}$ for PBC and as
$m=(C_{-}/4\sqrt{\pi})^{1/3}$ for APBC. In the case of APBC, owing
to the fact that $C_{-}<0$, the mass $m$ is given by a complex number
for a finite $L$. \\
\section{Gap equations on $S^{1} \times S^{1} \times S^{1}$}
\qquad In this section, we examine the gap equations on $S^{1} \times
S^{1} \times S^{1}$. For simplicity, we consider that the radii of the
circles are the same. We treat the case of APBC besides the case of PBC.
The first gap equation is $m^{2}b=0$ and we see $m=0$ or $b=0$.

The second gap equation under PBC is
\begin{eqnarray}
b^{2} &=& -{1 \over \rho^{3}}\lim_{s \rightarrow 1}\sum_{p,q,r}
\int_{0}^{\infty}dt{t^{s-1} \over \Gamma (s)}e^{-t({4\pi ^{2}
\over  \rho ^{2}}
(p^{2}+q^{2}+r^{2})+m^{2})} \nonumber \\
 &=& {1 \over 4\pi}(m+{6 \over \rho}\log(1-e^{-m\rho})-12\sum_{p,q}
{1 \over \sqrt{p^{2}+q^{2}}\rho}e^{-\sqrt{p^{2}+q^{2}}m\rho} \nonumber \\
& &-8\sum_{p,q,r}
{1 \over \sqrt{p^{2}+q^{2}+r^{2}}\rho}e^{-\sqrt{p^{2}+q^{2}+r^{2}}m\rho}),
\end{eqnarray}
where $\rho$ denotes the radius of $S^{1}$ and $p,q$ and $r$ run over
integers. If we take the limit $m \rightarrow 0$, the summations are
replaced by integrals and we get
\begin{equation}
b^{2}=\lim_{m \rightarrow 0}{1 \over 4\pi}(m+{6 \over \rho}\log m\rho
-{6\pi \over m\rho^{2}}-{4\pi \over m^{2}\rho^{3}})
\end{equation}
This equation can not hold. Thus we see $m\ne 0,b=0$.

The second gap equation under APBC is
\begin{eqnarray}
b^{2}&=& {1 \over 4\pi}(m+{2 \over \rho}\log(1-e^{-m\rho})
+{2\sqrt{2} \over \rho}\log(1-e^{-2m\rho})-\sum_{p,q}(4+8(-1)^{q})
{1 \over \sqrt{p^{2}+q^{2}}\rho}e^{-\sqrt{p^{2}+q^{2}}m\rho} \nonumber \\
& &-8\sum_{p,q,r}
(-1)^{r}
{1 \over \sqrt{p^{2}+q^{2}+r^{2}}\rho}e^{-\sqrt{p^{2}+q^{2}+r^{2}}m\rho}).
\end{eqnarray}
In the limit $m \rightarrow 0$, we have
\begin{equation}
b^{2}=\lim_{m \rightarrow 0}{1 \over 4\pi}(m+{2(1+\sqrt{2}) \over \rho}
\log m\rho-{2\sqrt{2} \over \rho}\log 2-{2\pi \over m\rho^{2}}).
\end{equation}
This equation can not hold. Thus we see $m\ne0,b=0$ under APBC.

We have seen that the scalar fields do not have a non-vanishing vacuum
expectation value on $S^{1} \times S^{1} \times S^{1}$ irrespective
of the boundary conditions. It will be not easy to see whether
corresponding solutions are stable or not.
\section{Summary}
\qquad In this paper, we have examined the gap equations of the
$O(N)$ non-linear
$\sigma$ model on three-dimensional manifolds of canstant curvature.
We  not only have considered imposing PBC, but also have cinsidered
imposing APBC in the
time direction to keep modular invariance of the partition function
in mind.  We have found difficulty
in choosing a solution on $S^{1} \times {\bf R}^{2}$ with APBC.
The two solutions both have good and bad points. One solution is
tachyonic and unstable.  The other solution is plausible in a sense,
but may conflict with the boundary condition itself. We have investigated
the gap equations on $S^{1} \times S^{2}$ in the two limiting cases.
The analysis based on $S^{1} \times {\bf R}^{2}$ suggests unstability
of a solution under APBC. Deviation from ${\bf R} \times S^{2}$
shows us how the solution becomes unstable under APBC. The boundary
condition restricts the spectrum of the laplacian, which brings about
the difference between PBC and APBC. Stability of a solution depends
on whether certain infinite summation over integers or half-integers
is larger than the corresponding definite integral or not. We have also
seen from the gap equations that spontaneous break down does not occur on
$S^{1} \times S^{1} \times S^{1}$ irrespective of boundary conditions.
\newpage
\begin{thebibliography} {99}
\bibitem{BPZ}A.A.Belavin,A.M.Polyakov and A.B.Zamolodchikov,Nucl.Phys.{\bf
B241}
(1984),333.
\bibitem{car1}J.L.Cardy, J.Phys.{\bf A18}(1985),L757.
\bibitem{are}I.Ya.Aref'eva, Theor.Mat.Fiz.{\bf 31}(1977),3;
I.Ya.Aref'eva and S.I.Azakov, Nucl.Phys.{\bf B162}(1980),298;
B.Rosenstein and B.J.Warr, Nucl.Phys.{\bf B336}(1990),435.
\bibitem{guru}S.Guruswamy, S.G.Rajeev and P.Vitale, Univ. Rochester Preprint
UR-1357(1994).
\bibitem{sach}S.Sachdev, Phys.Lett.{\bf B309}(1993),285.
\bibitem{fis}M.E.Fisher and P.-G. de Gennes, C.R.Acad.Sci.Ser.{\bf B287}(1978),
207;
A.H.Castro Neto and E.Fradkin, Nucl.Phys.{\bf B400}(1993),525.
\bibitem{cap}A.Cappeli, C.Itzykson and J.-B.Zuber, Nucl.Phys.{\bf
B280}(1987),445;
A.Kato, Mod.Phys.Lett.{\bf A2}(1987),585.
\bibitem{car2}J.L.Cardy, Nucl.Phys.{\bf B366}(1991),403.
\bibitem{fuji}A.Fujii and T.Inami, Kyoto Univ. Preprint YITP-K-1053(1994).
\bibitem{bir}N.D.Birrell and P.C.Davies, {\it Quantum Fields in Curved Space}
(Cambridge University Press, Cambridge, England, 1982).
\bibitem{car3}J.L.Cardy, Nucl.Phys.{\bf B290}(1987),355.
\end{document}